\documentclass[3p,sort&compress]{elsarticle}

\usepackage{amsmath,amssymb}
\usepackage{natbib}
\usepackage{graphicx}
\usepackage[utf8]{inputenc}
\newcommand{\coleq}{\mathrel{\mathop:}=}

\begin{document}

\begin{frontmatter}

\title{Complexity analysis of hypergeometric orthogonal polynomials}

\author[fm,cp]{J.S. Dehesa}
\ead{dehesa@ugr.es}
\author[fm,cp]{A. Guerrero}
\ead{agmartinez@ugr.es}
\author[ma,cp]{P. S\'anchez-Moreno}
\ead{pablos@ugr.es}

\address[fm]{Departamento de F\'{\i}sica At\'omica, Molecular y Nuclear, Universidad de Granada, Granada, Spain}
\address[ma]{Departamento de Matem\'atica Aplicada, Universidad de Granada, Granada, Spain}
\address[cp]{Instituto ``Carlos I'' de F\'{\i}sica Te\'orica y Computacional, Universidad de Granada, Granada, Spain}

\begin{abstract}
The complexity measures of the Cr\'amer-Rao, Fisher-Shannon and LMC (L\'opez-Ruiz, Mancini and Calvet) types of the
Rakhmanov probability density $\rho_n(x)=\omega(x) p_n^2(x)$ of the polynomials $p_n(x)$ orthogonal with respect to the weight function
 $\omega(x)$, $x\in (a,b)$, are used to quantify various two-fold facets of the spreading of the Hermite, Laguerre and Jacobi systems
all over their corresponding orthogonality intervals in both analytical and computational ways. Their explicit (Cr\'amer-Rao) and asymptotical 
(Fisher-Shannon, LMC) values are given for the three systems of orthogonal polynomials. Then, these complexity-type mathematical quantities are numerically examined in terms of the polynomial's degree $n$ and the parameters which characterize the weight function. 
Finally, several open problems about the generalised hypergeometric functions of Lauricella and Srivastava-Daoust types, as well as on the asymptotics of weighted $L_q$-norms of Laguerre and Jacobi polynomials are pointed out.
\end{abstract}

\end{frontmatter}

\section{Introduction}

The contents of this work are inspired to a great extent by the ideas of Leonhard Euler \cite{euler_1}, who long ago ushered in a true revolution in mathematics by combining painstaking observations (which he collected in 
notebooks) with use of quantities extracted from Physics (e.g., electrostatic properties of the zeros, capacity, mutual energy, logarithmic potential, ...) in order to gain further insights into the structure of mathematical functions, having encountered novel paths, notions and approaches which led to many of the
fundamental properties which presently we know about them. These Physics-based notions have contributed in a very important manner to the development of various mathematical fields, such as e.g. the theory of special functions, approximation theory and potential theory. With this spirit in mind, we use concepts and techniques extracted from Information Theory (such as entropy, entropic moments, complexity,...) to define, analyze and discuss new characteristics and structural properties of hypergeometric orthogonal polynomials.\\
The information-theoretical quantities of the Rakhmanov's probability density associated to these polynomials turn out to describe novel macroscopic facets of them, which possibly cannot be considered otherwise. Moreover, these quantities have not only a relevant mathematical character, but also they have an applied interest. Indeed, for example, they are closely related to physical entropies and measures of complexity of quantum systems which quantify their internal disorder and, consequently, they describe numerous fundamental and/or experimentally accesible quantities of these systems. This is essentially because the hypergeometric orthogonal polynomials often controls the mathematical description of the physical states of the quantum systems whose fundamental wave equations (Schrödinger, Dirac,...) are exactly- or quasi-exactly solvable.\\
A great challenge in contemporary science is to explore the mixing of simplicity and complexity, regularity and randomness, order and 
disorder, from particle physics and cosmology up to the adaptive complex systems and ultimately the living beings \cite{badii,gell_96,holland_12}. Many related
efforts have been spent in these and other disciplines in the last few decades, but still today not so many results have been
done in the theory of the mathematical functions which control the classical and quantum phenomena of the involved physical, chemical
and biological systems. Not even for the so-called special functions of mathematical physics and applied mathematics and, particularly,
the hypergeometric orthogonal polynomials \cite{nikiforov_1,temme_1}, which are very useful because of their numerous simple and elegant algebraic properties 
(e.g., recursion and ladder relations, Rodrigues formulas, integral representations, second-order differential equations). These 
``elementary'' functions have plenty of applications not only in many mathematical areas but also in applied sciences; in particular, they are used to model and interpret numerous scientific properties and phenomena, as well as to describe the wave functions of classical and quantum-mechanical states of a great deal of physical systems, beginning with the prototypic hydrogenic and oscillator-like systems \cite{nikiforov_1,galindo_3}.\\
The purpose of this paper is to quantify how simple or how complex are
the special functions of Applied Mathematics beginning by the classical or hypergeometric-type orthogonal polynomials
%the classical or hypergeometric-type orthogonal polynomials
in a real continuous variable (i.e., Hermite, Laguerre and Jacobi). The issues
\textit{How do we understand by simplicity and complexity?}  and  \textit{In what sense a certain mathematical function is simple and complex another
one?} are not at all simple. There does not exist a unique notion of complexity to grasp our intuitive notions in the appropriate 
manner. Or perhaps various different quantities (possibly not yet known) are required to grasp our intuitive notions of complexity of a mathematical function (e.g., a hypergeometric orthogonal polynomial) in order to capture the great diversity and complexity of its configuration shapes corresponding to the different values of its degree and the parameters which characterize its weight function. Up until now there does not exist such notions, to the best of our knowledge.\\
We should immediately say that various distinct complexity notions have been published in different contexts (dynamical systems, cellular automata, neuronal networks, social sciences, complex molecules, geophysical and astrophysical processes,...) for several purposes, such as e.g. to study pattern, structure and correlations in systems and processes. In addition, at times, some complexity-type quantities are successfully used to analyze the computational resources (space, time,...) required to solve a problem in computer science and quantum information theory \cite{goldreich_08,cubitt_14}, so at the interface of mathematics and computer science; more precisely, they concern the scaling of the resources in terms of the size of the problem. Nevertheless we will not use these complexity measures (heretoforth called by \textit{extrinsic} complexity measures) in the present work because they do depend on the context, such as e.g. the algorithmic and computational complexities; they are closely related to the time required for a computer to solve a given problem; so that it depends on the chosen computer. 

Here we will rather use density-dependent complexity measures, such as the Cram\'er-Rao, Fisher-Shannon and LMC (L\'opez-Ruiz-Mancini-Calvet)
complexities, recently introduced in a quantum-physical context (see e.g. the reviews \cite{angulo_1,dehesa_3} and \cite{martin_06,dehesa_09,dehesa_09bis,molina_12,lopezrosa_jmp13}), which are of \textit{intrinsic} character in the sense that they do not depend on the context but on the quantum probability density of the system under consideration. Our goal is to quantify how simple or how
complex are the classical orthogonal polynomials $p_n(x)$  by means of the complexity measures of its associated Rakhmanov's probability density
\cite{rakhmanov_1}. Remark that, contrary to other complexity notions (algorithmic, computational,...)\cite{goldreich_08,cubitt_14},
the density-dependent complexities are intrinsic properties of the polynomials. Thus, the intrinsic complexity notions are closely
related to the main macroscopic features of the associated probability density of the polynomials (irregularities, extent, fluctuations, smoothing,...).

The structure of the paper is the following. In Section 2 we define and describe the meaning of the complexity measures of the classical
orthogonal polynomials which we use throughout the paper. Then, in Section 3 we give the values of the Cram\'er-Rao complexity of the Hermite 
polynomials and the asymptotics (infinity) of the Fisher-Shannon and LMC complexities of Hermite polynomials. In Sections 4 and 5 we find
the Cram\'er-Rao complexity and the disequilibrium as well as the asymptotics of the Fisher-Shannon and LMC complexities of the Laguerre and 
Jacobi polynomials, respectively. In Section 6, the previous analytical results are numerically discussed in terms of the polynomial's degree and the parameters which characterize the weight function. Finally, some conclusions and
various open problems found throughout the paper are given.

\section{Complexity measures of a general probability density}

In this Section we give the definitions and mathematical meanings of the complexity measures of a probability distribution.

Let us consider a general one-dimensional random variable $X$ characterized by the continuous probability distribution $\rho(x)$,
$x \in \Lambda \subseteq \mathbb{R}$. To quantify the spread of $X$ over the interval $\Lambda$ we usually employ the statistical root-mean-square
or standard deviation $\Delta x$, which is the square root of the variance
   \begin{equation*}
      V[\rho]=\left(\Delta x\right)^\frac12=\langle x^2 \rangle-\langle x \rangle^2,
   \end{equation*}
where
\[
\langle f(x) \rangle= \int_{\Lambda} f(x) \rho(x) dx.
\]
   
The information theory provides other spreading measures such as the R\'enyi and Shannon entropies and the Fisher information. The R\'enyi
entropy $R_q[\rho]$ of $\rho(x)$ is defined \cite{renyi_70} by
   \begin{equation*}
       R_q[\rho]\coleq \frac{1}{1-q}\ln W_q[\rho]=\frac{1}{1-q}\ln \int_{\Lambda} [\rho(x)]^q dx,
   \end{equation*}
where $W_q[\rho]=\langle \rho^{q-1} \rangle$ denotes the $q$th-order frequency or entropic moment of $\rho(x)$. The limiting value $q\rightarrow
1$, taking into account the normalization condition $W_1[\rho]=1$, yields the Shannon entropy \cite{shannon_49}
  \begin{equation*}
     S[\rho]\coleq \lim_{q\rightarrow 1} R_q[\rho]=-\int_{\Lambda} \rho(x) \ln \rho(x) dx.
  \end{equation*}

The Fisher information of $\rho(x)$ is defined \cite{frieden_04,fisher_pcps25} as
  \begin{equation*}
     F\left[\rho\right]:=\int_{\Lambda} \frac{\left(\frac{d}{dx} \rho(x)\right)^2}{\rho(x)}\,dx.
  \end{equation*}

It is worth remarking that: (a) these three information-theoretic spreading measures do not depend on any particular point of their
interval $\Lambda$, contrary to the standard deviation, (b) the Fisher information has a locality property because it is a functional 
of the derivative of $\rho(x)$, and (c) the standard deviation and the R\'enyi and Shannon entropies are global properties because
they are power and logarithmic functionals of $\rho(x)$, respectively. Moreover they have different units, so that they can not be compared each
other. To overcome this difficulty, the following information-theoretic lengths have been introduced \cite{hall_pra99}
  \[
    N_q[\rho]=\exp \left(R_q[\rho]\right)=\left(W_q[\rho]\right)^{\frac{1}{1-q}}, \quad {\rm R\acute{e}nyi\; length},
  \]
  \begin{equation*}
      N_1[\rho]=\lim_{q\rightarrow 1}  N_q[\rho]=\exp\left(S[\rho]\right), \quad {\rm Shannon\; length},
  \end{equation*}
  \[
    \delta x=\frac{1}{\sqrt{F[\rho]}}, \quad \quad {\rm Fisher\; length}.
  \]
It is straightforward to observe that these three lengths, as well as the standard deviation $\Delta x$, have the same units of $X$.

Let us highlight that the quantities ($V[\rho]$, $R_q[\rho]$, $S[\rho]$, $F[\rho]$), and its related measures
($\Delta x$, $N_q[\rho]$, $N_1[\rho]$, $\delta x$), are complementary since each of them grasps a single different facet of the
probability density $\rho(x)$. So, the variance measures the concentration of the density around the centroid while the R\'enyi and
Shannon entropies are measures of the extent to which the density is in fact concentrated, and the Fisher information is a quantitative estimation of the 
oscillatory character of the density since it measures the pointwise concentration of the probability over its support interval $\Lambda$.

Recently, some composite density-dependent information-theoretic quantities have been introduced; namely, the complexity measures of Cr\'amer-Rao \cite{dembo_itit91,dehesa_1,antolin_ijqc09},
Fisher-Shannon \cite{angulo_pla08,romera_1} and L\'opez-Ruiz-Mancini-Calbet (LMC) \cite{catalan_pre02} types. They are given by the product of two of the previous single spreading measures
as
  \begin{equation}
      \label{cramerrao}
      C_{CR}[\rho]=F[\rho] \times V[\rho],
  \end{equation}
  \begin{equation}
     \label{fishershannon}
      C_{FS}[\rho]=F[\rho] \times \frac{1}{2 \pi e} e^{2 S[\rho]}=\frac{1}{2 \pi e} F[\rho] \times N_1^2[\rho],
  \end{equation}
  \begin{equation}
     \label{lmc}
     C_{LMC}[\rho]=W_2[\rho] \times e^{S[\rho]}=\langle \rho \rangle \times N_1[\rho],
  \end{equation}
for the Cr\'amer-Rao, Fisher-Shannon  and LMC complexities, respectively. Each of them grasps the combined balance of two different facets
of the probability density. The Cr\'amer-Rao complexity quantifies the wiggliness or gradient content of $\rho(x)$ jointly with the probability
spreading around the centroid. The Fisher-Shannon complexity measures the gradient content of $\rho(x)$ together with its total extent 
in the support interval. The LMC complexity measures the combined balance of the average height of $\rho(x)$ (as given by the second-order
entropic moment $W_2[\rho]$, also called disequilibrium $D[\rho]$), and its total extent (as given by the Shannon entropic power
$N[\rho]=e^{S[\rho]}$).\\
Moreover, it may be easily observed that these three complexity measures are (a) dimensionless, (b) bounded from
below by unity (when $\rho$ is a continuous density in $\mathbb{R}$ in the Cr\'amer-Rao and Fisher-Shannon cases, and for any $\rho$ in the LMC case), and (c) minimum for the two extreme (or least complex) distributions which correspond to perfect order (i.e. the extremely
localized Dirac delta distribution) and maximum disorder (associated to a highly flat distribution). Finally, they fulfil invariance properties
under replication, translation and scaling transformation \cite{yamano_jmp04,yamano_pa04}.

\section{Complexity measures of Hermite polynomials}

In this section we give the values of the Cr\'amer-Rao complexity, as well as the asymptotics ($n \rightarrow \infty$) of
the Fisher-Shannon and LMC complexities, of the Hermite polynomials $H_n(x)$ characterized by the orthogonality condition (see e.g. \cite{sanchezmoreno_2,temme_1})
   \[
     \int_{-\infty}^{+\infty} H_n(x) H_m(x) e^{-x^2} dx=\delta_{m,n}, \quad m,n\in \mathbb{N}.
   \]
These quantities are defined by the corresponding complexity measures of the Rakhmanov-Hermite probability density
   \[
    \rho_H(x)=H_n^2(x) e^{-x^2}.
   \]

Let us begin with the Cr\'amer-Rao complexity which, according to Eq.(\ref{cramerrao}), is given by
   \[
    C_{CR}[\rho_H]=F[\rho_H] \times V[\rho_H],
   \]
where the variance and the Fisher information of the Hermite polynomials are known \cite{dehesa_1,sanchezruiz_1} to be
   \[
     V[\rho_H]=n+\frac{1}{2},
   \]
and
  \begin{equation}
     \label{varfis_hermite}
     F[\rho_H]=4n+2,
  \end{equation}
respectively. Therefore, one easily has the value
  \begin{equation*}
     C_{CR}[\rho_H]=4n^2+4n+1,
  \end{equation*}
for the Cr\'amer-Rao quantity.

Similarly, from Eq. (\ref{fishershannon}) one has that the Fisher-Shannon complexity of Hermite polynomials is given by
   \[
     C_{FS}[\rho_H]=F[\rho_H] \times \frac{1}{2 \pi e} N_1^2[\rho_H],
   \]
where the Shannon length (also called Shannon entropy power) of the Hermite polynomials, $N_1[\rho_H]=\exp(S[\rho_H])$, have not been analytically calculated up until now except in the
asymptotical case \cite{sanchezmoreno_1}:
  \begin{equation}
     \label{shannonlength_asy}
     N_1[\rho_H]\approx \frac{\pi}{e} \sqrt{2n}; \quad n \gg 1.
  \end{equation}
Then, this expression together with the Fisher value (\ref{varfis_hermite}) directly lead to the asymptotical value of the Fisher-Shannon
of the Hermite polynomials:
  \begin{equation*}
    C_{FS}\left[\rho_H\right]\approx \left(\frac{4 \pi}{e^3}\right) n^{2}, \qquad n\gg 1.
  \end{equation*}

Finally, from Eq. (\ref{lmc}) one obtains the LMC complexity of Hermite polynomials as
   \[
    C_{LMC}[\rho_H]=W_2[\rho_H] \times N_1[\rho_H],
   \]
where the second-order entropic moment (also called disequilibrium)
   \[
     W_2[\rho_H]=\langle \rho_H \rangle,
   \]
can be explicitly calculated both for all $n$ and in the asymptotic case. The latter value is
   \[
     W_2[\rho_H]= 2\pi^{-2} (2n)^{-\frac12}\left(\ln(n)+O(1)\right); n\gg 1,
   \]
as explained in \cite{aptekarev_12}. Then, this expression together with Eq. (\ref{shannonlength_asy}) gives
   \[
     C_{LMC}[\rho_H]\approx    \frac{2}{\pi e}\ln n; \qquad n\gg 1,
   \]
for the asymptotical value of the LMC complexity of the Hermite polynomials $H_n(x)$.

\section{Complexity measures of Laguerre polynomials}

In this section we give the values of the Cr\'amer-Rao complexity and the asymptotical value of the Fisher-Shannon of the Laguerre
polynomials $L_n^{(\alpha)}(x),\alpha>-1$. As well, we point out the issues to calculate the LMC complexity of these mathematical objects both
in the general (i.e., for all $n$) and asymptotical (i.e., at large $n$) cases. The Rakhmanov probability density associated to the Laguerre
polynomials $L_n^{(\alpha)}(x)$ characterized by the orthogonality condition (see e.g. \cite{temme_1,olver_1})
   \[
     \int_0^{+\infty} L_n^{(\alpha)}(x) L_m^{(\alpha)}(x) x^{\alpha} e^{-x} dx=\delta_{mn},
   \]
is defined by
   \[
     \rho_L(x)=\left[L_n^{(\alpha)}(x)\right]^2 x^{\alpha} e^{-x}.
   \]

Then, according to Eq.(\ref{cramerrao}), the Cr\'amer-Rao complexity of the Laguerre polynomials is given by
   \begin{equation}
      \label{cramerrao_l}
      C_{CR}[\rho_L]=F[\rho_L] \times V[\rho_L],
   \end{equation}
where the variance and the Fisher information are given \cite{dehesa_1,sanchezruiz_1} by
   \begin{equation}
       \label{variance_l}
       V[\rho_L]=2 n^2+2(\alpha+1)n+\alpha+1,
   \end{equation}
and
   \begin{eqnarray}
       \label{fisher_l}
       F[\rho_L]=
       \left\{
       \begin{array}{ll}
          4n+1, & \alpha=0, \\
          \frac{(2n+1)\alpha+1}{\alpha^2-1}, & \alpha>1, \\
          \infty , & \alpha \in [-1,+1],\alpha \neq 0,
       \end{array}
       \right.
   \end{eqnarray}
respectively. The expressions (\ref{cramerrao_l})-(\ref{fisher_l}) lead to the following value
   \begin{eqnarray*}
        C_{CR}[\rho_L] =
        \left\{
        \begin{array}{ll}
           8n^3+\left[8(\alpha+1)+2\right]n^2+6(\alpha+1)n+(\alpha+1),& \alpha=0, \\ & \\
           \frac{1}{\alpha^2-1}\left[4\alpha n^3+(4 \alpha^2+6\alpha+2)n^2+(4 \alpha^2+6\alpha+2)n+(\alpha+1)^2\right], & \alpha>1, \\ & \\
           \infty, & \hspace{-1cm}{\rm otherwise,}
        \end{array}
        \right.
   \end{eqnarray*}
for the Cr\'amer-Rao complexity of Laguerre polynomials.

Let us now consider the Fisher-Shannon complexity of these polynomials which is
defined, according to Eq.(\ref{fishershannon}), by
   \begin{equation}
      \label{fishershannon_l}
      C_{FS}[\rho_L]=F[\rho_L] \times \frac{1}{2 \pi e} N_1^2[\rho_L],
   \end{equation}
where the Shannon length or Shannon entropy power $N_1[\rho_L]=\exp(S[\rho_L])$ of the Laguerre polynomial $L_n^{(\alpha)}(x)$ is not yet
known for all values of the degree $n$, mainly because it is a logarithmic functional of the polynomial. However, its asymptotical (large $n$)
value has been found \cite{sanchezmoreno_2} to be
   \begin{equation}
      \label{shannon_length}
      N_1[\rho_L] \approx \frac{2 \pi n}{e}.
   \end{equation}
Then, from Eqs.(\ref{fisher_l}), (\ref{fishershannon_l}) and (\ref{shannon_length}) one obtains the following asymptotics for the Fisher-Shannon
complexity of the Laguerre polynomial $L_n^{(\alpha)}(x)$:
   \begin{eqnarray*}
      C_{FS}\left[\rho_L\right]\approx
      \left\{
      \begin{array}{ll}
      \left(\frac{8 \pi}{e^3}\right) n^{3},& \alpha=0, \\ & \\
      \frac{4 \alpha}{\alpha^2-1} \left(\frac{\pi}{e^3}\right) n^{3}, & \alpha>1, \\ & \\
      \infty, & {\rm otherwise}.
      \end{array}
      \right.
   \end{eqnarray*}

Finally, let us tackle the calculation of the LMC complexity of Laguerre polynomials which is given by
   \begin{equation*}
      C_{LMC}[\rho_L]=W_2[\rho_L] \times N_1[\rho_L].
   \end{equation*}
Now, we have two opposite situations when calculating these two factors: while the Shannon length $N_1[\rho_L]$ is only
known in the asymptotics case (see Eq.\ref{shannon_length}), the second-order entropic moment $W_2[\rho_L]$ has been recently shown 
\cite{sanchezmoreno_2} to be expressed in the two following manners for all values of the degree $n$:

    \begin{itemize}
      \item[(i)] In terms of the four-variate Lauricella function $F_A^{(4)}\left(\frac12,\frac12,\frac12,\frac12\right)$ \cite{srivastava_85}:
        \begin{eqnarray}
            \label{disequilibrium_laguerre}
            W_2[\rho_L] &=&\left( \frac{n!}{\Gamma(\alpha+n+1)} \right)^2
            \frac{\Gamma(2\alpha+1)}{2^{2\alpha+1}}
            \left(
            \begin{array}{c}
            n+\alpha\\
            n
            \end{array}
            \right)^{4}\nonumber\\
             &&\times F_A^{(4)}
             \left( 
             \begin{array}{c}
	     2\alpha+1; -n,-n,-n,-n\\
	     \alpha+1,\alpha+1,\alpha+1,\alpha+1
             \end{array}
             ;\frac{1}{2},\frac{1}{2},\frac{1}{2},\frac{1}{2}
             \right).
         \end{eqnarray}

      \item[(ii)] In terms of the multivariate Bell polynomials $B_{m,l}(a_1,a_2\ldots,a_{m-l+1})$ \cite{guerrero_1}:
         \begin{equation}
            \label{disequilibrium_lag_bell}
             W_2[\rho_L] =\left[ \sum_{k=0}^{4n}\frac{\Gamma(2\alpha+k+1)}
            {2^{2\alpha+k+1}} \frac{(4)!}{(k+4)!}
            B_{k+4,4}\left(c_0^{(n,\alpha)},2!c_1^{(n,\alpha)},...,(k+1)!c_k^{(n,\alpha)}\right)\right],
         \end{equation}
       where the parameters $c_t^{(n,\alpha)}$ are given by
          \[
            c_t^{(n,\alpha)}=\sqrt{\frac{\Gamma(n+\alpha+1)}{n!}}\frac{(-1)^t}{\Gamma(\alpha+t+1)}
            \left(
            \begin{array}{c}
            n\\
            t
            \end{array}
            \right).
          \]
    \end{itemize}

Taking into account that $N_1[\rho_L]$ is not known for a generic degree $n$ of the polynomials and the asymptotics of 
Eqs.(\ref{disequilibrium_laguerre}) and (\ref{disequilibrium_lag_bell}) is a formidable task, the evaluation of the LMC complexity
of Laguerre polynomials remains to be an open problem in both general and asymptotic situations of $n$.

\section{Complexity measures of Jacobi polynomials}

In this section we give the values of the Cr\'amer-Rao complexity of the Jacobi polynomials $P_n^{(\alpha,\beta)}(x)$, with
$\alpha,\beta>-1$, as well as the asymptotics (large $n$) of its Fisher-Shannon complexity. In addition, we discuss the reasons
why the evaluation of the LMC complexity cannot yet be done. The Jacobi polynomials are well-known to satisfy the orthogonality
condition (see \cite{olver_1,temme_1})
   \[
     \int_{-1}^{+1} P_n^{(\alpha,\beta)}(x) P_m^{(\alpha,\beta)}(x) (1-x)^{\alpha} (1+x)^{\beta} dx=\delta_{mn},
   \]
and its associated Rakhmanov probability density $\rho_J(x)$ is given by
   \[
     \rho_J(x) =\left[P_n^{(\alpha,\beta)}(x)\right]^2 (1-x)^{\alpha} (1+x)^{\beta}.
   \]

The Cr\'amer-Rao complexity of $P_n^{(\alpha,\beta)}(x)$ is defined by the Cr\'amer-Rao of the density $\rho_J(x)$ which, according
to Eq.(\ref{cramerrao}), is given by
   \begin{equation}
       \label{cramerrao_jacobi}
       C_{CR}[\rho_J]=F[\rho_J] \times V[\rho_J].
   \end{equation}
These two factors have been recently calculated  \cite{dehesa_1,sanchezruiz_1}, having the values
   \begin{eqnarray}
     V[\rho_J]=
     \frac{4(n+1)(n+\alpha+1)(n+\beta+1)(n+\alpha+\beta+1)}{(2n+\alpha+\beta+1)(2n+\alpha+\beta+2)^2 (2n+\alpha+\beta+3)}\nonumber\\
     +\frac{4n(n+\alpha)(n+\beta)(n+\alpha+\beta)}{(2n+\alpha+\beta-1)(2n+\alpha+\beta)^2 (2n+\alpha+\beta+1)},
     \label{eq_standard_deviation_jacobi}
   \end{eqnarray}
for the variance, and
    \begin{eqnarray}
       \label{fisher_jacobi}
        F\left[\rho_J\right]=
       \left\{
       \begin{array}{ll}
       2n(n+1)(2n+1), & \alpha,\beta=0, \\[3mm]
       \frac{2n+\beta+1}{4}\left[\frac{n^2}{\beta+1}+n+(4n+1)(n+\beta+1)+\frac{(n+1)^2}{\beta-1}\right],
       & \alpha=0, \beta>1, \\[3mm]
       \frac{2n+\alpha+\beta+1}{4(n+\alpha+\beta-1)}\left[n(n+\alpha+\beta-1)
       \left(\frac{n+\alpha}{\beta+1}+2+\frac{n+\beta}{\alpha+1}\right)\right.& \\
       +\left.(n+1)(n+\alpha+\beta)\left(\frac{n+\alpha}{\beta-1}+2+\frac{n+\beta}{\alpha-1}\right)\right],
       &\alpha, \beta>1, \\[3mm]
       \infty, & {\rm otherwise,}
       \end{array}
       \right.
    \end{eqnarray}
for the Fisher information of the Jacobi polynomial $P_n^{(\alpha,\beta)}(x)$. Then, from 
Eqs.(\ref{cramerrao_jacobi})-(\ref{fisher_jacobi}) one obtains the value
    \begin{eqnarray*}
       C_{CR}\left[\rho_J\right]=
      \left\{
      \begin{array}{ll}
      2n(n+1)\left[\frac{(n+1)^2}{2n+3}+\frac{n^2}{2n-1}\right],& \alpha=\beta=0, \\ & \\
      \left[\frac{(n+1)^2(n+\beta+1)^2}{(2n+\beta+2)^2(2n+\beta+3)}+\frac{n^2(n+\beta)^2}{(2n+\beta-1)(2n+\beta)^2}\right] & \\
      \times \left[\frac{n^2}{\beta+1}+n+(4n+1)(n+\beta+1)+\frac{(n+1)^2}{\beta-1}\right], & \alpha=0,\beta>1, \\ & \\
      \left[\frac{(n+1)(n+\alpha+1)(n+\beta+1)(n+\alpha+\beta+1)}{(2n+\alpha+\beta+2)^2(2n+\alpha+\beta+3)}+
      \frac{n(n+\alpha)(n+\beta)(n+\alpha+\beta)}{(2n+\alpha+\beta-1)(2n+\alpha+\beta)^2}\right] & \\
      \times \frac{1}{n+\alpha+\beta-1}\left[n(n+\alpha+\beta-1)\left(\frac{n+\alpha}{\beta+1}+2+\frac{n+\beta}{\alpha+1}\right)\right. & \\
      \left.+(n+1)(n+\alpha+\beta)\left(\frac{n+\alpha}{\beta-1}+2+\frac{n+\beta}{\alpha-1}\right)\right], & \alpha>1,\beta>1, \\ & \\
      \infty, & {\rm otherwise}.
      \end{array}
      \right.
     \end{eqnarray*}

The Fisher-Shannon complexity of Jacobi polynomial is, according to Eq.(\ref{fishershannon}), given by
     \begin{equation}
        \label{fishershannon_jacobi}
        C_{FS}[\rho_J]=F[\rho_J] \times \frac{1}{2 \pi e} N_1^2[\rho_J].
     \end{equation}
We cannot calculate the exact value of this complexity measure for all values of the polynomial degree $n$ since the Shannon length
$N_1[\rho_J]$ has not yet been found for a generic $n$, because of its logarithmic-functional nature. However, its asymptotic value has been
recently shown \cite{guerrero_1} to be as
     \begin{equation}
        \label{shannon_length_jacobi}
        N_1[\rho_J] \approx \frac{\pi}{e}, \quad n\rightarrow \infty,
     \end{equation}
so that the asymptotics of the Fisher-Shannon complexity of the Jacobi polynomial $P_n^{(\alpha,\beta)}(x)$ is
    \begin{eqnarray*}
      C_{FS}\left[\rho_J\right]\approx
      \left\{
      \begin{array}{ll}
      \left(\frac{2 \pi}{e^3}\right) n^{3},& \alpha=\beta=0, \\ & \\
      \frac{1}{4}\left(\frac{\pi}{e^3}\right) \left[\frac{1}{\beta+1}+4+\frac{1}{\beta-1}\right] n^3, & \alpha=0,\beta>,1 \\ & \\
      \frac{1}{2}\left(\frac{\pi}{e^3}\right) \left[\frac{\beta}{\beta^2-1}+\frac{\alpha}{\alpha^2-1}\right] n^3, & \alpha>1,\beta>1, \\ & \\
      \infty, & {\rm otherwise}.
      \end{array}
      \right.
    \end{eqnarray*}
where Eqs.(\ref{fisher_jacobi}), (\ref{fishershannon_jacobi}) and (\ref{shannon_length_jacobi}) have been taken into account.

Finally it is worth highlightening that the LMC complexity of Jacobi polynomial defined by
   \begin{equation*}
       C_{LMC}[\rho_J]=W_2[\rho_J] \times N_1[\rho_J],
   \end{equation*}
cannot yet be evaluated neither for a generic polynomial degree $n$, nor in the asymptotic case $n \rightarrow \infty$. This is so despite
we know \cite{sanchezmoreno_12,guerrero_1} the asymptotic behavior of the Shannon length $N_1[\rho_J]$ and the two following expressions for the second-order entropic
moment $W_2[\rho_J]$ (also called disequilibrium):

    \begin{itemize}
      \item[(a)] In terms of the four-variate Srivastava-Daoust function $F^{1:2;2;2;2}_{1:1;1;1;1}(1,1,1,1)$ \cite{sanchezmoreno_12} \cite{srivastava_85}:
         \begin{equation*}
            W_2[\rho_J]=D\left[P_n^{(\alpha,\beta)}\right]=\frac{d_0^{(2\alpha,2\beta)}}{\left( d_n^{(\alpha,\beta)} \right)^2}
            b_0\left( 4,n,\alpha,\beta,2\alpha,2\beta \right),
         \end{equation*}
       where
         \[
           d_n^{(\alpha,\beta)}=
           \frac{2^{\alpha+\beta+1} \Gamma(\alpha+n+1) \Gamma(\beta+n+1)}{n!(\alpha+\beta+2n+1)\Gamma(\alpha+\beta+n+1)},
         \]
       and
         \begin{multline*}
            b_0(4,n,\alpha,\beta,2\alpha,2\beta) =
            \left(
            \begin{array}{c}
            n+\alpha\\
            n
            \end{array}
            \right)^{4}\\
             \times
            F^{1:2;2;2;2}_{1:1;1;1;1}
            \left(
            \begin{array}{c}
            2\alpha+1: -n,\alpha+\beta+n+1;\ldots; -n,\alpha+\beta+n+1\\
            2\alpha+2\beta+2: \alpha+1; \ldots;\alpha+1
            \end{array}
            ; 1,1,1,1
            \right).
         \end{multline*}

       \item[(b)] In terms of the Bell polynomials $B_{m,l}(a_1,a_2,\ldots,a_{m-l+1})$ \cite{guerrero_1}:
           \begin{equation*}
              W_2[\rho_J]=\sum_{k=0}^{4n} \frac{(4)!}{(k+4)!}
              B_{k+4,4}\left(c_0^{(n,\alpha,\beta)},2!c_1^{(n,\alpha,\beta)},...,(k+1)!c_k^{(n,\alpha,\beta)}\right)
              \mathcal{I}(k,2,\alpha,\beta),
           \end{equation*}
         where the coefficients $c_{t}^{(n,\alpha,\beta)}$ are given by
           \begin{eqnarray*}
             c_{t}^{(n,\alpha,\beta)}&=&\sqrt{\frac{\Gamma(\alpha +n+1)(2n+\alpha +\beta +1)}{n!2^{\alpha +\beta +1}\Gamma(\alpha +\beta +n+1)\Gamma(n+\beta +1)}}\\
	             &&\times\sum_{i=t}^{n}(-1)^{i-t}
	             \left(
	             \begin{array}{c}
	             n\\i
	             \end{array}
	             \right)
	             	             \left(
	             	             \begin{array}{c}
	             	             i\\t
	             	             \end{array}
	             	             \right)
             \frac{\Gamma(\alpha +\beta +n+i+1)}{2^{i}\Gamma(\alpha +i+1)},
           \end{eqnarray*}
         and
           \[
              \mathcal{I}(k,q,\alpha,\beta) =
             \frac{(-1)^k 2^{1+\alpha q+\beta q}\Gamma(\alpha q+1)\Gamma(\beta q+1)}{\Gamma(\alpha q+\beta q+2)}
             \,_2F_1 \left(
             \begin{array}{l}
             -k,1+\beta q\\
              2+(\alpha+\beta)q
             \end{array}
             ;2\right).
           \]
           
          To find the value of the LMC complexity of $P_n^{(\alpha,\beta)}(x)$ we would further need to know the explicit value
          of the Shannon length $N_1[\rho]$ and/or the asymptotics of the disequilibrium $D[P_n^{(\alpha,\beta)}(x)]$, what is a
          formidable task. Therefore the analytical knowledge of the LMC complexity of Jacobi polynomials remains to be an open 
          problem in both general and asymptotic cases.
    \end{itemize}

\section{Numerical discussion}

In this section the expressions for the Cramer-Rao, Fisher-Shannon and LMC complexities found in the three previous sections are numerically studied for the three classical families of orthogonal polynomials of Hermite, Laguerre and Jacobi.
The values of these complexities are computationally discussed in terms of the degree and parameters of the corresponding polynomials.

\subsection{Dependence on the polynomial's degree $n$}

\begin{figure}
\begin{center}
\includegraphics[width=8cm]{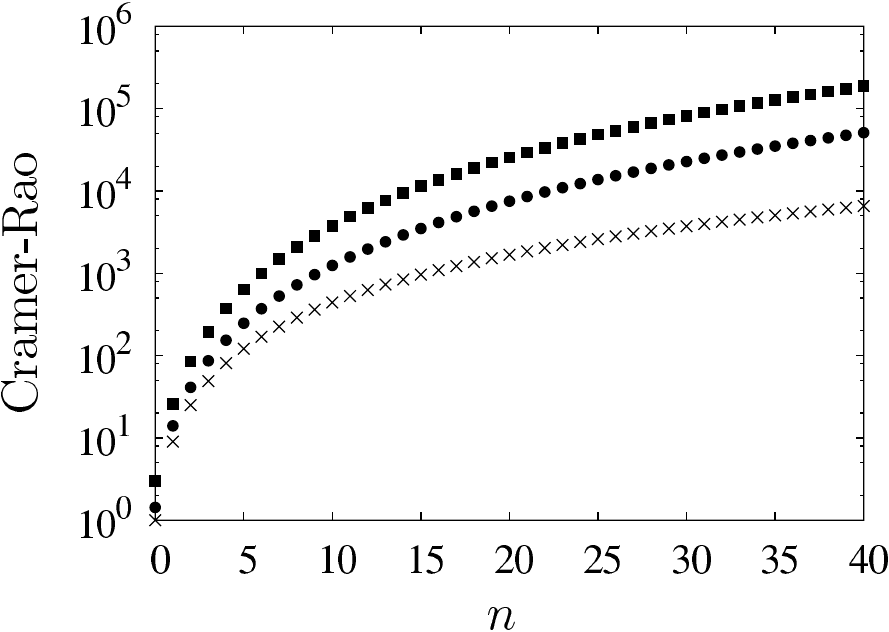}
\end{center}
\caption{Cramer-Rao complexity measure for the Rakhmanov densities of the Hermite $H_n(x)$ ($\times$), Laguerre $L_n^{(2)}(x)$ ($\blacksquare$) and Jacobi $P_n^{(2,2)}(x)$ ($\bullet$) polynomials as a function of the degree $n$ for $n=0,1,\ldots,40$.}
\label{fig1}
\end{figure}

\begin{figure}
\begin{center}
\includegraphics[width=8cm]{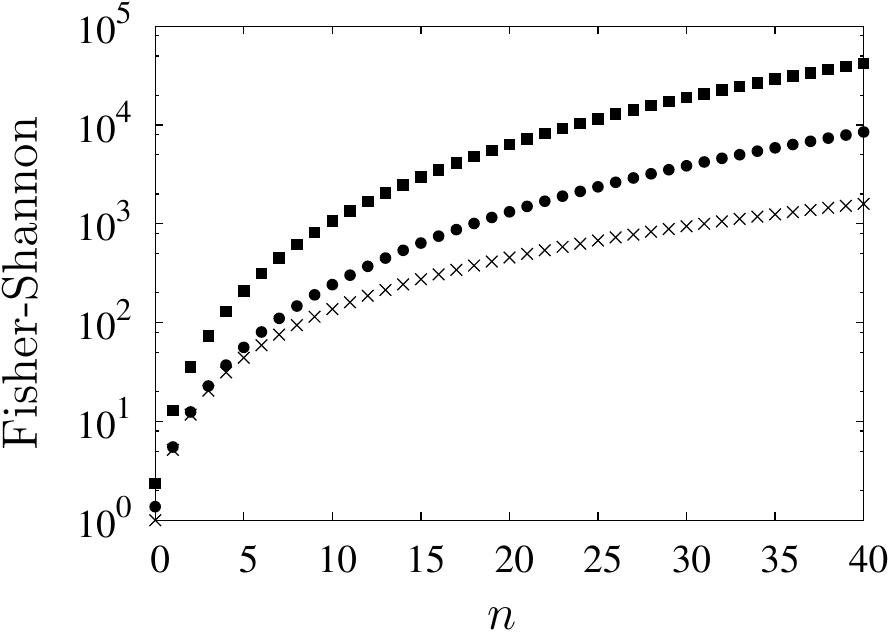}
\end{center}
\caption{Fisher-Shannon complexity measure for  the Rakhmanov densities of the Hermite $H_n(x)$ ($\times$), Laguerre $L_n^{(2)}(x)$ ($\blacksquare$) and Jacobi $P_n^{(2,2)}(x)$ ($\bullet$) polynomials as a function of the degree $n$ for $n=0,1,\ldots,40$.}
\label{fig2}
\end{figure}

Figures \ref{fig1} and \ref{fig2} represent the Cramer-Rao and the Fisher-Shannon complexity measures, respectively, for the Rakhmanov densities of the Hermite $H_n(x)$ ($\times$), Laguerre $L_n^{(2)}(x)$ ($\blacksquare$) and Jacobi $P_n^{(2,2)}(x)$ ($\bullet$) polynomials as a function of the degree $n$ for $n=0,1,\ldots,40$. In the three cases these complexity measures monotonically grow with the degree $n$.
For different values of the parameters of the Laguerre and Jacobi polynomials, the behaviour of these measures is the same.
This behaviour can be explained from an intuitive idea of complexity: The number of maxima of the Rakhmanov density associated to a polynomial of degree $n$ is equal to $n+1$, with one zero between each two consecutive maxima. Therefore, the number of oscillations and, consequently, one form of complexity of the density increases with $n$. Then, it looks like an intuitive idea of complexity is in agreement with the values of these complexity measures.

\begin{figure}
\begin{center}
\includegraphics[width=8cm]{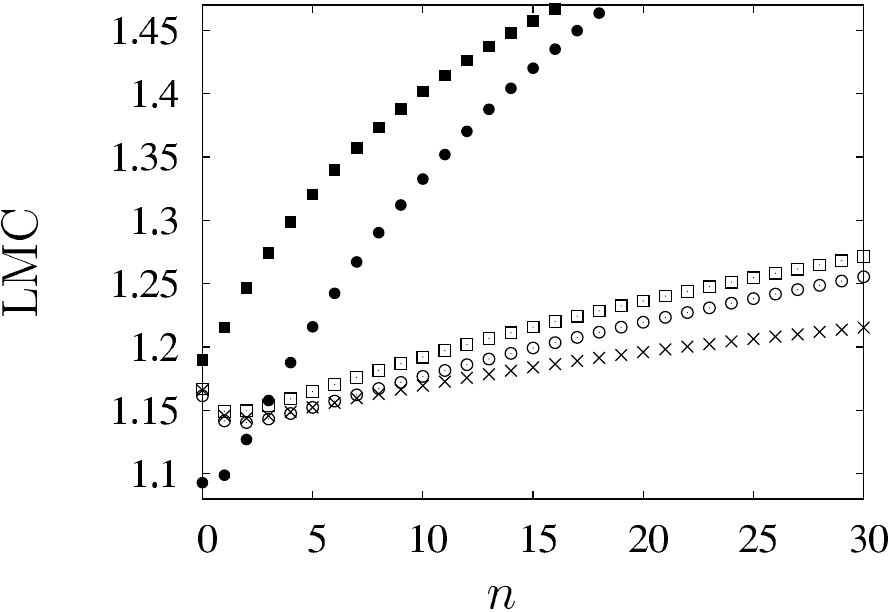}
\end{center}
\caption{LMC complexity measure for the Rakhmanov densities of the Hermite $H_n(x)$ ($\times$), Laguerre $L_n^{(2)}(x)$ ($\blacksquare$) and $L_n^{(50)}(x)$ ($\square$), and Jacobi $P_n^{(2,2)}(x)$ ($\bullet$) and $P_n^{(50,50)}(x)$ ($\circ$) polynomials, as a function of the degree $n$ for $n=0,1,\ldots,30$.}
\label{fig3}
\end{figure}

Figure \ref{fig3} shows the LMC complexity measure for the Rakhmanov densities of the Hermite $H_n(x)$ ($\times$), Laguerre $L_n^{(2)}(x)$ ($\blacksquare$) and $L_n^{(50)}(x)$ ($\square$), and Jacobi $P_n^{(2,2)}(x)$ ($\bullet$) and $P_n^{(50,50)}(x)$ ($\circ$) polynomials, as a function of the degree $n$ for $n=0,1,\ldots,30$.
This complexity measure is a monotonically increasing function of $n$ in the Laguerre and Jacobi cases with small values of the parameters, as can be seen in the figure for the polynomials $L_n^{(2)}(x)$ and $P_n^{(2,2)}(x)$.
However, the LMC complexity measure is a decreasing function of $n$ for small values of $n$ in the Hermite case and also for Laguerre and Jacobi polynomials with large values of the parameters ($\alpha=\beta=50$ in the figure).
This can be explained taking into account that the LMC complexity does not depend on the Fisher information or any other information measure sensitive to the oscillatory character of the density. Then, the increase of the oscillatory content of the density from one maximum ($n=0$) to two maxima ($n=1$) is not very relevant to the increment of the complexity, from the point of view of the LMC measure. What provokes the decreasing of the LMC complexity is the  decrement experienced by the average height of the density, that is measured by the disequilibrium; while the Shannon entropy remains almost constant. Thus, the LMC complexity is much more sensitive to the smoothness of the density than to its oscillatory character, grasping a different intuitive idea of complexity.

\subsection{Dependence on the polynomial's parameters}

\begin{figure}
\begin{center}
\includegraphics[width=8cm]{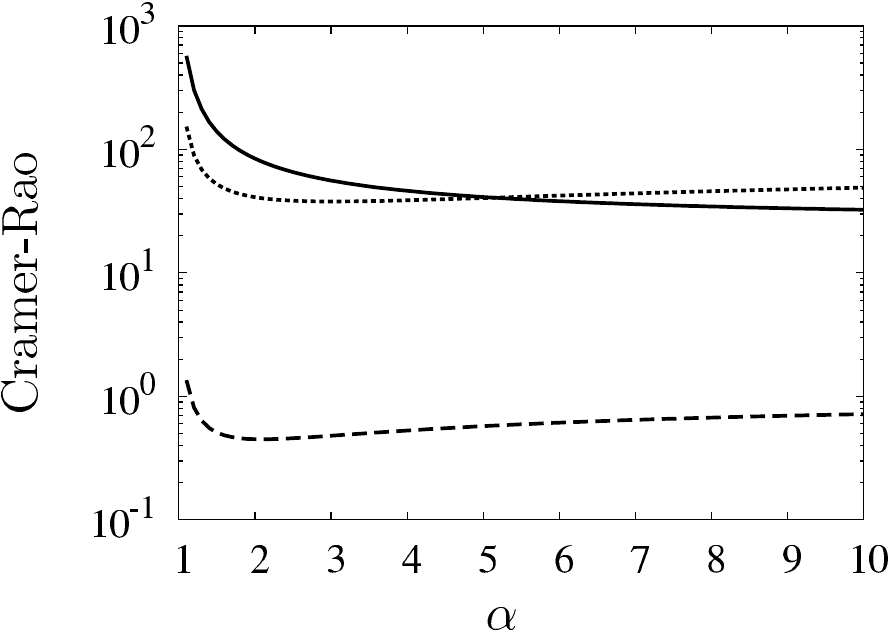}
\end{center}
\caption{Cramer-Rao complexity measure for the Rakhmanov densities of the Laguerre $L_2^{(\alpha)}(x)$ (solid line), and the Jacobi $P_0^{(\alpha,0)}(x)$ (dashed line) and $P_2^{(\alpha,2)}(x)$ (dotted line) polynomials, as a function of the parameter $\alpha$, for $1<\alpha <10$.}
\label{fig4}
\end{figure}

\begin{figure}
\begin{center}
\includegraphics[width=8cm]{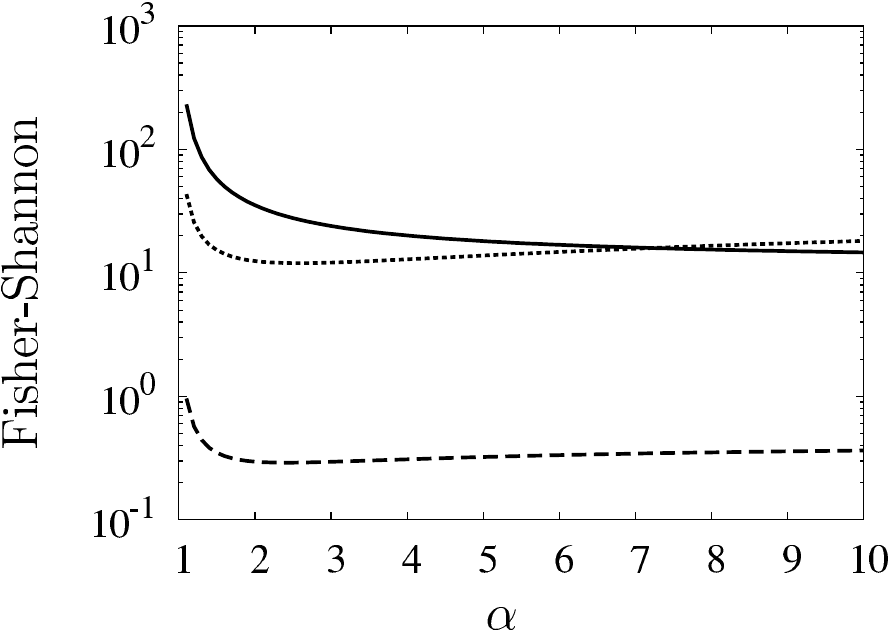}
\end{center}
\caption{Fisher-Shannon complexity measure for the Rakhmanov densities of the Laguerre $L_2^{(\alpha)}(x)$ (solid line), and the Jacobi $P_0^{(\alpha,0)}(x)$ (dashed line) and $P_2^{(\alpha,2)}(x)$ (dotted line) polynomials, as a function of the parameter $\alpha$, for $1<\alpha <10$.}
\label{fig5}
\end{figure}

\begin{figure}
\begin{center}
\includegraphics[width=8cm]{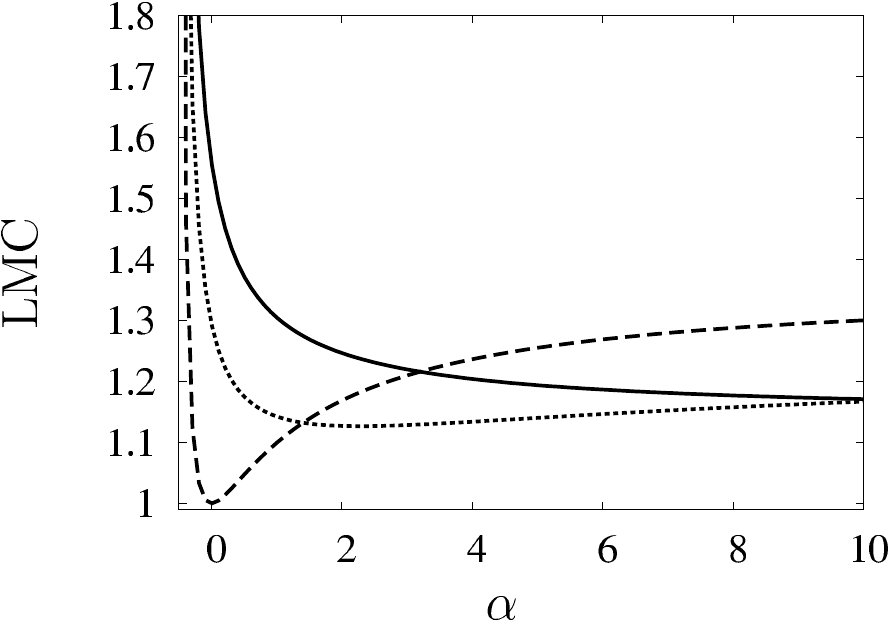}
\end{center}
\caption{LMC complexity measure for the Rakhmanov densities of the Laguerre $L_2^{(\alpha)}(x)$ (solid line), and the Jacobi $P_0^{(\alpha,0)}(x)$ (dashed line) and $P_2^{(\alpha,2)}(x)$ (dotted line) polynomials, as a function of the parameter $\alpha$, for $-\frac12<\alpha <10$.}
\label{fig6}
\end{figure}

Figures \ref{fig4}, \ref{fig5} and \ref{fig6} represent the Cramer-Rao, Fisher-Shannon and LMC complexity measures, respectively, for the Rakhmanov densities of the Laguerre $L_2^{(\alpha)}(x)$ (solid line), and the Jacobi $P_0^{(\alpha,0)}(x)$ (dashed line) and $P_2^{(\alpha,2)}(x)$ (dotted line) polynomials, as a function of the parameter $\alpha$, for $1<\alpha <10$ in the Cramer-Rao and Fisher-Shannon case (as the Fisher information is defined for $\alpha>1$, apart from the discrete value $\alpha=0$), and $-\frac12 <\alpha<10$ in the LMC case (as the disequilibrium is defined for $\alpha>-\frac12$).

For the Laguerre polynomials, the three complexity measures decrease when $\alpha$ increases.
Again, this behaviour can be explained from an intuitive idea of complexity: The spreading of these densities increases with $\alpha$. Since the complexity decreases (and the smoothness increases) as the spreading grows, the complexity of these densities decreases as $\alpha$ increases, as shown by the three studied complexity measures.

In the Jacobi case, the principal characteristic on these figures is the minima that show the three complexity measures as a function of $\alpha$.
This behaviour can also be explained from an intuitive point of view: For a given value of $\beta\in \{0\}\cap(1,+\infty)$, if $\alpha\le 1$ the density appears more concentrated around positive values of $x$ in the interval $(-1,1)$. However, as $\alpha$ increases from this value the density moves to negative values of $x$. Within this transition, the density pass through a configuration of maximal spreading and smoothness, or minimal complexity, that is detected by these complexity measures with the minima that we can see in these figures.
Another feature from these figures is the clear separation between points for $n=0$ and points for $n=2$ that appear in the Cramer-Rao and Fisher-Shannon representations, contrary to the LMC complexity measure. This behaviour is due to the effect of the Fisher information, very sensitive to the oscillatory character, in the first two complexity measures. However, the LMC complexity measure depend on the disequilibrium, that is not affected directly by the oscillatory content but for the average height of the density. For the three measures, since the density is defined in a bounded interval, the variance and the Shannon entropy have an upper bound, so the variations of these measures come essentially from the Fisher information and the disequilibrium, respectively.

\section{Conclusions and open problems}

Nowadays the concept of complexity has become fundamental in Science and Technology because of its usefulness to interpret, explain and predict numerous natural phenomena. However its mathematical realization is manifold, depending not only on the specific discipline where it was created but also on the concrete purpose which generated it. Generally speaking, the different notions of complexity published in the literature can be classified as intrinsic (i.e., the ones which depend on the single-particle probability density of the many-body system under consideration) and extrinsic (i.e., the ones which do not depend on any probability density and take into account the context where the system is related with). The complexities of extrinsic character (e.g., Kolmogorov, computational and algorithmic complexities) were earlier introduced, being mostly used in technological areas \cite{goldreich_08,cubitt_14}, while the intrinsic ones (e.g., Cramer-Rao, Fisher-Shannon and LMC complexities) have been recently introduced, being mostly used in scientific disciplines to discuss the internal structure of physical systems as well as to describe the course of chemical and biological processes and reactions \cite{angulo_1,dehesa_3,dehesa_09,dehesa_09bis,antolin_ijqc09,angulo_pla08}.

In this work we have introduced various complexities of intrinsic character to study the complexity of the hypergeometric-type orthogonal polynomials in a real continuous variable. We have defined them as the corresponding complexities of the Rakhmanov probability density associated to these polynomials. Then, we have discussed both algebraically and numerically the complexities of Hermite, Laguerre and Jacobi polynomials in terms of both the polynomial degree and the characterizing parameters of their weight functions. Analytically, we have found the explicit expression of the Cramer-Rao complexity in the Hermite case and the asymptotics of the Fisher-Shannon and LMC complexities in the three Hermite, Laguerre and Jacobi cases. Numerically we have shown that, opposite to the single information-theoretic measures (e.g., Shannon entropy, Fisher information, disequilibrium,...), these three composite complexities grasp different aspects of the complex nature that people have about the mathematical functions here considered.

In addition, several open problems have been pointed out throughout the paper.
Here we would like to highlight two important issues in the field of generalized hypergeometric functions (namely, the reduction of the Lauricella function $F_A^{(4)}\left(\frac12,\frac12,\frac12,\frac12\right)$ and the Srivastava-Daoust function $F^{1:2;2;2;2}_{1:1;1;1;1}(1,1,1,1)$ to much simpler functions) and the following asymptotic problem of classical orthogonal polynomials $y_n(x)$: to find the asymptotical ($n\to\infty$) value of the functional
\[
W_q[\rho]=\int_\Lambda \left[\rho(x)\right]^q dx = \int_\Lambda \left[\omega(x)\right]^q \left|y_n(x)\right|^{2q}dx,
\]
which is very closely connected to the weighted $L_q$-norm of these polynomials. Here, $\omega(x)$ ($x\in\Lambda$) is the weight function with respect to which these polynomials are orthogonal, and $\rho(x)=\omega(x) |y_n(x)|^2$ denotes the associated Rakhmanov probability density. This asymptotical issue has been recently solved for the Hermite polynomials \cite{aptekarev_12} but it remains open in the Laguerre and Jacobi cases. The solution of these issues would allow one to calculate not only the disequilibrium but also the LMC complexity of the Laguerre and Jacobi cases and, \emph{in extenso}, the corresponding quantities of numerous physical systems whose quantum-mechanical states are described by wavefunctions controlled by these polynomials. Finally, for the sake of completeness, let us also comment that the asymptotics ($q\to\infty$) of these mathematical objects have been recently considered \cite{dehesa_jmc14}.

\section*{Acknowledgements}
This work has been partially funded by the Junta-de-Andaluc\'{\i}a grants FQM-207, FQM-7276 and FQM-4643 as well as the MICINN grant FIS2011-24540.

\bibliographystyle{model1-num-names}
\bibliography{complexity_polynomials}

\end{document}